\documentclass[aps,amssymb,amsmath,prl,twocolumn, showpacs, superscriptaddress]{revtex4}
\usepackage{graphicx}
\usepackage{dcolumn}
\usepackage{bm}

\usepackage{amssymb,amsmath,amsfonts,latexsym,graphicx,subfigure,verbatim}
\usepackage[matrix,frame,arrow]{xy}
\usepackage{epsf}

\newcommand{\ket}[1]{| #1 \rangle}
\newcommand{\bra}[1]{\langle #1 |}

\begin{document}
\title{Tripartite entanglement versus tripartite nonlocality in 3-qubit GHZ-class states}

\author{S. Ghose}
\affiliation{Department of Physics and Computer Science, Wilfrid Laurier University, Waterloo Ontario N2L 3C5, Canada}
\author{N. Sinclair}
\affiliation{Department of Physics and Computer Science, Wilfrid Laurier University, Waterloo Ontario N2L 3C5, Canada}
\affiliation{Institute for Quantum Information Science, University of Calgary, Calgary, Alberta T2N 1N4, Canada}
\author{S. Debnath}
\affiliation{Department of Physics and Computer Science, Wilfrid Laurier University, Waterloo Ontario N2L 3C5, Canada}
\affiliation{Indian Institute of Technology, Powai, Mumbai 400076, India}
\author{P. Rungta}
\affiliation{NMR Research Centre, 
Indian Institute of Science, Bangalore - 560012, India.}
\author{R. Stock}
\affiliation{Department of Physics, University of Toronto, Toronto, Ontario M5S 1A7, Canada}

\date{\today}


\begin{abstract}
We analyze the relationship between tripartite entanglement and genuine tripartite nonlocality for 3-qubit pure states in the GHZ class. We consider a family of states known as the generalized GHZ states and derive an analytical expression relating the 3-tangle, which quantifies tripartite entanglement, to the Svetlichny inequality, which is a Bell-type inequality that is violated only when all three qubits are nonlocally correlated. We show that states  with 3-tangle less than 1/2 do not violate the Svetlichny inequality. On the other hand, a set of states known as the maximal slice states do violate the Svetlichny inequality, and exactly analogous to the two-qubit case, the amount of violation is directly related to the degree of tripartite entanglement. We discuss further interesting properties of the generalized GHZ and maximal slice states.
\end{abstract}


\maketitle

{\it Introduction}: 
Quantum theory allows correlations between spatially separated systems that are fundamentally different from classical correlations. This difference becomes evident when entangled states violate Bell-type inequalities~\cite{Bell} that place an upper bound on the correlations compatible with local hidden variable (or local realistic) theories. All pure entangled states of two qubits violate the Bell-type Clauser-Horner-Shimony-Holt (CHSH) inequality~\cite{CHSH}, and the amount of violation increases with the degree of bipartite entanglement~\cite{Gisin1, Popescu1} in the state. In this paper, we generalize this 2-qubit relationship to important 3-qubit pure states in the GHZ class~\cite{GHZClass}. We find analytical expressions relating tripartite entanglement to a Bell-type inequality formulated by Svetlichny~\cite{Svetlichny} that tests for tripartite nonlocal correlations, and we identify unique nonlocal properties of certain states.  
Our work is motivated not only by foundational implications~\cite{Gisin2}, but also by recent theoretical and experimental interest in multiqubit entanglement and nonlocality for  
novel applications  in quantum communication and quantum computation~\cite{Theory, Review, Seevinck, Expts, GHZPrep}.  
Nonlocal correlations of three or more particles may also play an integral role in phase transitions and criticality in many body systems~\cite{Review}. Furthermore, our analysis allows the possibility of generalization to N particles, which would provide new avenues for the understanding of many-body condensed matter, optical and atomic systems.

The study of  Bell inequalities for 3-qubit states is complicated by the problem of distinguishing between violations arising from 2-qubit versus 3-qubit correlations~\cite{NQubit, Cereceda}. We focus here on the Svetlichny inequality, because its violation is a sufficient condition for the confirmation of genuine 3-qubit nonlocal correlations~\cite{Svetlichny}.
We identify and discuss special nonlocal properties of two subsets of the GHZ class~\cite{GHZClass}: the generalized GHZ (GGHZ) states $\ket{\psi_g}$ and the maximal slice (MS) states~$\ket{\psi_s}$~\cite{Slice},
\begin{eqnarray}
\ket{\psi_g} &=& \cos\theta_1 |000\rangle+\sin\theta_1 |111\rangle \label{gghz}\;,\\
\ket{\psi_s} &=&\frac{1}{\sqrt{2}}\Bigl\{\ket{000}+\ket{11}\{\cos\theta_3\ket{0}+\sin\theta_3\ket{1}\}\Bigr\}\;.\label{slice}
\end{eqnarray}
These states have unique entanglement properties due their inherent symmetries~\cite{Slice}, which makes them interesting candidates for information processing protocols. For instance, the well-known GHZ state, common to both subsets ($\theta_1= \pi/4, \theta_3=\pi/2)$, has been prepared in different physical systems and is a resource for various practical applications~\cite{GHZPrep}.

Like other Bell-type inequalities, the Svetlichny inequality is defined in terms of the expectation value of a Bell-type operator $S$ that is bounded by the inequality $|\langle S\rangle|\leq 4$~\cite{Svetlichny}. We show that the maximum expectation value of $S$ for the GGHZ and MS states is
\begin{eqnarray}
S_{\text{max}}(\psi_g) &=&
\left\{
\begin{array}{lr}
4 \sqrt{1-\tau(\psi_g)}\;, & \tau(\psi_g) \leq 1/3 \\
4\sqrt{2\tau(\psi_g)}\;, & \tau(\psi_g) \geq 1/3\;,
\end{array}
\right.
\label{gghzresult0}\\
S_{\text{max}}(\psi_s)&=&4\sqrt{1+\tau(\psi_s)},\label{stauslice}
\end{eqnarray}
where the  {\it 3-tangle}  $\tau(\psi)$  quantifies tripartite entanglement~\cite{3tangle}, with $\tau(\psi_g)=\sin^2 2\theta_1$ and $\tau(\psi_s)=\sin^2 \theta_3$. 
Our results reveal interesting properties of the GGHZ and MS states. For the GGHZ states, $S_{\text{max}}(\psi_g)$ initially decreases monotonically with $\tau$, and then increases for $\tau>1/3$. The Svetlichny inequality is only violated by GGHZ states with  $\tau > 1/2$. However, all MS states violate the Svetlichny inequality and Eq.~(\ref{stauslice}) is exactly analogous to the well-known $2$-qubit relationship between bipartite entanglement and the CHSH inequality~\cite{Gisin1, Popescu1}.
Our analysis shows that within a particular 3-parameter family that is experimentally accessible, the MS states achieve the maximum possible value of   $S_{\text{max}}$ for a given  $\tau$, while conversely, the GGHZ states yield the minimum possible $S_{\text{max}}$. Our expressions also provide a practical way to measure the tripartite entanglement $\tau$ via measurement of $S_{\text{max}}$, which involves only local measurements of each qubit.  
 
{\it The 3-tangle}: In order to facilitate the discussion of our results, we first briefly describe the $3$-tangle $\tau$, a measure of genuine tripartite entanglement~\cite{3tangle} defined as
\begin{equation}
\tau = {\cal C}_{1(23)}^2-{\cal C}_{12}^2-{\cal C}_{13}^2.
\end{equation} 
${\cal C}^2_{1(23)}$  measures the entanglement between qubit $1$ and the joint state of qubits 2 and 3. 
The concurrences ${\cal C}_{12}$ and ${\cal C}_{13}$ quantify bipartite  entanglement between qubits $1$ and $2$ and $1$ and $3$ respectively (for further details see~\cite{Wootters}). 
The 3-tangle is invariant under permutation of the indices $1$, $2$ and $3$ and is bounded between 0 (for separable states) and 1 (for the maximally entangled GHZ state).  For GGHZ states,  ${\cal C}_{12} = {\cal C}_{13} = 0$ and hence, 
$\tau(\psi_g) = {\cal C}_{1(23)}^2(\psi_g)=\sin^2{2\theta_1}$.
For MS states, ${\cal C}_{1(23)}=1$, ${\cal C}_{12}= \cos^2\theta_3$ and ${\cal C}_{13} = 0$. So   
$\tau(\psi_s) =\sin^2{\theta_3}$.

{\it Svetlichny's Inequality}: Bell-type inequalities based on absolute local realism, where all three qubits are locally but realistically correlated, fails to distinguish between bipartite and tripartite nonlocality~\cite{NQubit}. For instance, Mermin's inequality~\cite{Mermin} is violated by biseparable states in which two of the qubits are separable from the third~\cite{NQubit, Cereceda}, and hence it cannot unambiguously identify genuine tripartite nonlocality. 
Svetlichny therefore considered a hybrid model of nonlocal-local realism~\cite{Svetlichny} where two of the qubits are nonlocally correlated, but are locally correlated to the third. 
Suppose we have an ensemble of three spatially separated qubits, and the measurements $A={\vec{a}} \cdot {\vec{\sigma_1}}$ or $A^\prime={\vec{a}^\prime} \cdot {\vec{\sigma_1}}$ are performed on qubit $1$, $B={\vec{b}} \cdot {\vec{\sigma_2}}\cdot$ or $B^\prime={\vec{b}^\prime} \cdot{\vec{\sigma_2}}$ on qubit $2$, and $C={\vec{c}}\cdot {\vec{\sigma_3}}$ or $C^\prime={\vec{c}^{\;\prime}}\cdot{\vec{\sigma_3}}$ on qubit 3, where $\vec{a}, \vec{a}', \vec{b},\vec{b}'$ and $\vec{c}, \vec{c}'$ are unit vectors and the $\vec{\sigma}_i$ are spin projection operators that can be written in terms of the  Pauli matrices. 
The Svetlichny operator is defined as
\begin{eqnarray}
S &=& A(BK+B'K')+A'(BK'-B'K),  
\label{soperator}
\end{eqnarray}
where $K=C+C'$ and $K'=C-C'$.
If a theory is consistent with a hybrid model of nonlocal-local realism, then the expectation value for any $3$-qubit state is bounded by Svetlichny's inequality, $|\langle\Psi|S|\Psi\rangle| \equiv S(\Psi)\leq 4$,  which is maximally violated by the GHZ state~\cite{Svetlichny}.
By design, all biseparable states satisfy the Svetlichny inequality. Therefore it is only violated when all three qubits are nonlocally correlated.  

In order to find the maximum expectation value of $S$ for the 3-qubit GGHZ states and MS states, we adapt the technique used to derive the 2-qubit result~\cite{Popescu1}.  Let ${\vec{a}}=(\sin\theta_a\cos\phi_a, \sin\theta_a\sin\phi_a,\cos\theta_a)$,
and likewise define ${\vec{a}^\prime}$ ${\vec{b}}$, ${\vec{b}^\prime}$, ${\vec{c}}$ and ${\vec{c}^\prime}$. In addition, define unit vectors $\vec{d}$ and $\vec{d}'$ such that ${\vec{b}}+{\vec{b^\prime}}=2{\vec{d}}\cos\theta$ and ${\vec{b}}-{\vec{b^\prime}}=2{\vec{d^\prime}}\sin\theta$. 
Thus
\begin{equation}
{\vec{d}}\cdot{\vec{d^\prime}}=\cos\theta_d\cos\theta_{d^\prime}+\sin\theta_d\sin\theta_{d^\prime}\cos(\phi_d - \phi_{d'})=0.
\label{orthogonal}
\end{equation}
Then setting $D=\vec{d}\cdot \vec{\sigma}_2$ and $D'=\vec{d'}\cdot \vec{\sigma}_2$, the expectation value of $S$ (Eq.~(\ref{soperator})) for a state $|\Psi \rangle$ can be rewritten as
\begin{eqnarray}
S(\Psi)&=& 2|\cos\theta\langle ADC \rangle+ \sin\theta\langle AD'C'\rangle\nonumber\\
&&+\; \sin\theta\langle A'D'C\rangle-
\cos\theta\langle A'DC'\rangle|\nonumber\\
& \leq &  2\Big | \Big\{\langle ADC\rangle^2
+\langle AD'C' \rangle^2\Big\}^{1\over 2}\nonumber\\
&&+\; \Big\{\langle A'D'C\rangle^2+
\langle A'DC' \rangle^2\Big\}^{1\over 2} \Big |,
\label{inequality0}
\end{eqnarray}
where we have used the fact that
\begin{equation}
\label{maxtheta}
x\cos\theta+y\sin\theta\leq (x^2+y^2)^{1\over 2}\;,
\end{equation}
with the equality holding when $\tan\theta=y/x$. All square roots are taken to be positive. 
We now use Eq.~(\ref{inequality0}) to obtain the main results of the paper.

{\it The GGHZ States:} The first term in Eq.~(\ref{inequality0}) with respect to the GGHZ states gives
\begin{eqnarray}
\bra{\psi_g} ADC\ket{\psi_g}
&=&\cos{2\theta_1}\textrm{cos}\theta_a\textrm{cos}\theta_d\textrm{cos}\theta_c\nonumber\\
&&+\; \sin{2\theta_1}\textrm{sin}\theta_a\textrm{sin}\theta_d\textrm{sin}\theta_c\textrm{cos}\phi_{adc}\nonumber\\
&\leq & \Big\{\cos^2{2\theta_1}\cos^2\theta_a\cos^2\theta_d\nonumber\\
&&+\;\sin^2{2\theta_1}\sin^2\theta_a\sin^2\theta_d\Big\}^{1\over 2}
\label{inequality1}
\end{eqnarray}
where we have applied Eq.~(\ref{maxtheta}) with respect to $\theta_c$, and chosen $\cos^2\phi_{adc}\equiv\cos^2(\phi_a+\phi_d+\phi_c)=1$. Then equations (\ref{inequality0}) and (\ref{inequality1}) imply
\begin{widetext}
\begin{eqnarray}
S(\psi_g) &\leq & 2\Big\{\cos^2{2\theta_1}(\cos^2\theta_d+\cos^2\theta_{d^\prime})\cos^2\theta_a
+\sin^2{2\theta_1}(\sin^2\theta_d+\sin^2\theta_{d^\prime})\sin^2\theta_a\Big\}^{1\over 2}\nonumber\\
&+& 2\Big\{\cos^2{2\theta_1}(\cos^2\theta_d+\cos^2\theta_{d^\prime})\cos^2\theta_{a^\prime}
+\sin^2{2\theta_1}(\sin^2\theta_d+\sin^2\theta_{d^\prime})\sin^2\theta_{a^\prime}\Big\}^{1\over 2},
\end{eqnarray}
which when maximized with respect to $\theta_a$ and $\theta_{a^\prime}$,  gives 
\begin{equation}
S(\psi_g)
\leq
\left\{
\begin{array}{lr}
4\cos{2\theta_1}(\cos^2\theta_d+\cos^2\theta_{d^\prime})^{1\over 2}\;, & \cos^2{2\theta_1} (\cos^2\theta_d+\cos^2\theta_{d^\prime})\geq\sin^2{2\theta_1} (\sin^2\theta_d+\sin^2\theta_{d^\prime})\\
4\sin{2\theta_1}(\sin^2\theta_d+\sin^2\theta_{d^\prime})^{1\over 2}\;, & \cos^2{2\theta_1} (\cos^2\theta_d+\cos^2\theta_{d^\prime})\leq\sin^2{2\theta_1} (\sin^2\theta_d+\sin^2\theta_{d^\prime}).
\end{array}
\right.
\label{inequality3}
\end{equation}
\end{widetext} 
Here we have used the fact that
\begin{eqnarray}
\label{maxtheta2}
x\sin^2\theta+y\cos^2\theta 
\leq \left \{
\begin{array}{lr} 
y, \; & x \leq y\;\\
x, \;& x \geq y\;
\end{array}
\right. \;,
\end{eqnarray}
with the first inequality realized when $\theta=0$ or $\pi$ and the second  when $\theta=\pi/2$.
Now using Eq.~(\ref{orthogonal}), the maximum of $\cos^2\theta_d+\cos^2\theta_{d^\prime}$ is 1 while the maximum of $\sin^2\theta_d+\sin^2\theta_{d^\prime}$ is 2. 
Inserting these values into Eq.~(\ref{inequality3}) and using $\tau(\psi_g)=\sin^2{2\theta_1}$ yields the form of Eq.~(\ref{gghzresult0}),
\begin{equation}
\label{inequality5}
S(\psi_g)\leq
\left\{
\begin{array}{lr}
4 \sqrt{1-\tau(\psi_g)}, & \tau(\psi_g) \leq 1/3 \\
4\sqrt{2\tau(\psi_g)}\;, & \tau(\psi_g) \geq 1/3\;.
\end{array}
\right.
\end{equation}
The equality in Eq.~(\ref{inequality5}), $S_{\text{max}}(\psi_g)$, is realized by the following possible sets of unit vectors: for $\tau\leq {1/3}$, ${\vec{a}}$, ${\vec{a}^\prime}$, ${\vec{b}}$, ${\vec{b}^\prime}$ and ${\vec{c}}$  are all aligned along ${\vec{z}}$, and  ${\vec{c}^\prime}$ is aligned along ${-\vec{z}}$; for $\tau\geq {1/3}$, all the measurement vectors lie in the $xy$-plane with $\phi_{adc}=\phi_{ad'c'}=\phi_{a'd'c}=0$, $\phi_{a'dc'}=\pi$ and $\phi_d - \phi_{d'}=\pi/2$.
This change in the measurement direction at $\tau=1/3$ produces a sharp change in $S_{\text{max}}(\psi_g)$ as illustrated in Fig. $1$: as $\tau$ is increased from $0$ to ${1/3}$, $S_{\text{max}}(\psi_g)$ actually decreases, after which $S_{\text{max}}(\psi_g)$ monotonically increases with $\tau$. When $\tau\leq {1/2}$, GGHZ states do not violate Svetlichny's inequality. 
Notice that the nonviolation however, does not prevent us from experimentally measuring the entanglement of GGHZ states.  To do so, we can choose for example, the unit vectors in the $xy$-plane  identified earlier and experimentally measure the expectation value of $S(\psi_g)$. For these measurement angles,
$S(\psi_g)=4\sqrt{2\tau(\psi_g)}$, from which we can compute the entanglement $\tau(\psi_g)$. 
In the regime $\tau\leq {1/3}$, the measured value of $S(\psi_g)=4\sqrt{2\tau(\psi_g)}$ is not the maximum possible value, but this is not important if the goal is only to measure the entanglement $\tau(\psi_g)$.
\begin{figure}
\includegraphics[width=0.25\textwidth]{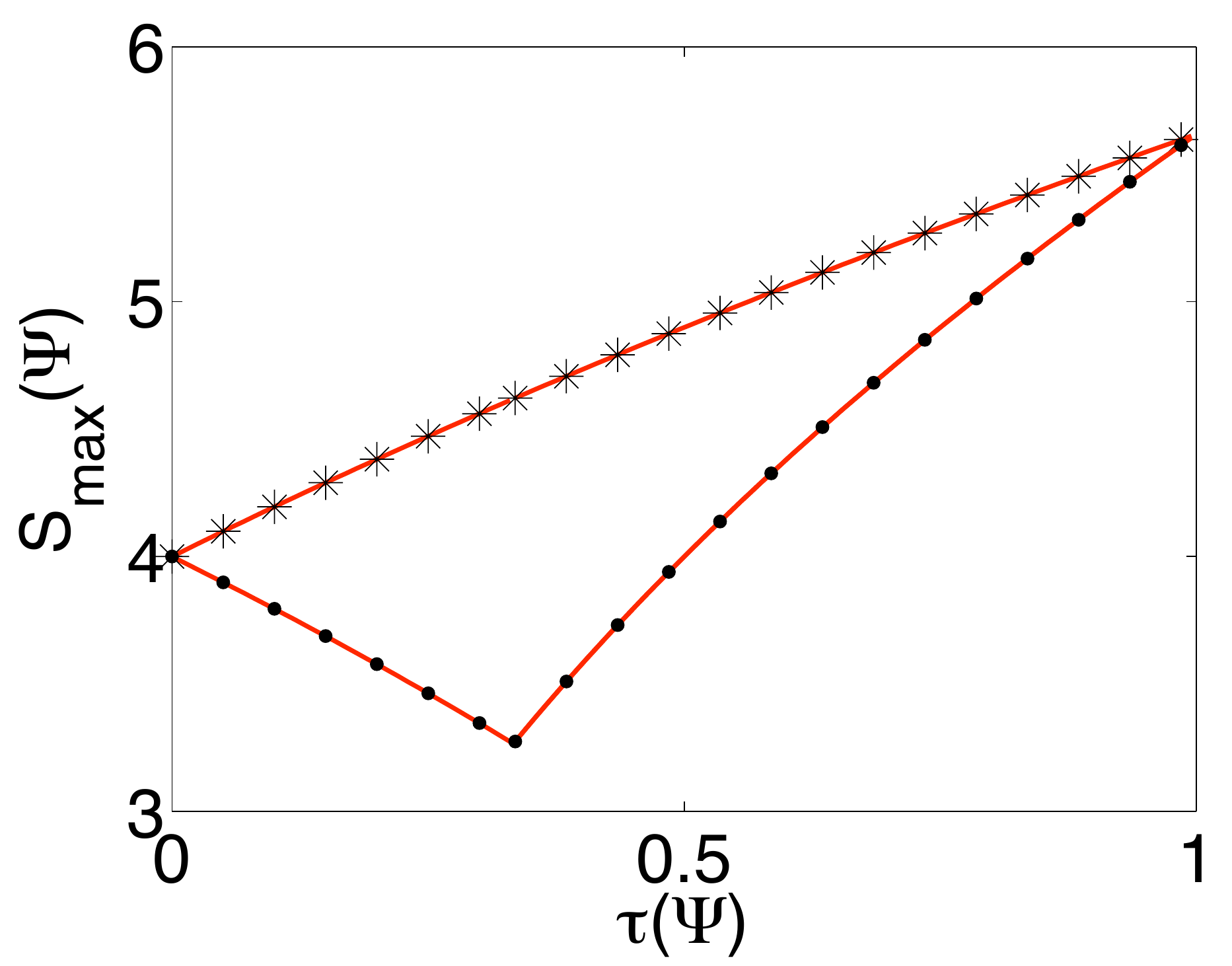}
\caption{ 
Dots show a plot of  Eq.~(\ref{gghzresult0}) for $S_{\text{max}}(\psi_g)$  versus $\tau$ for the GGHZ states. Comparison to the numerical bounds~\cite{Emary}  of Eq.~(\ref{emmabound}) (solid) shows agreement with the lower bound. Stars show a plot  of  Eq.~(\ref{stauslice}) for $S_{\text{max}}(\psi_s)$  versus $\tau$ for the MS states. Comparison to the numerical bounds~\cite{Emary} of Eq.~(\ref{emmabound}) (solid) shows agreement with the upper bound.}
\end{figure}

The GGHZ states belong to the 3-parameter family,
\begin{eqnarray}
\label{3parameter}
\ket{\psi_3}&=&\text{cos}\theta_1\ket{000}+\text{sin}\theta_1\ket{\phi_1 \phi_2 \phi_3},
\end{eqnarray}
where 
$\ket{\phi_1}=\ket{1}$, $\ket{\phi_2}=\text{cos}\theta_2\ket{0}+\text{sin}\theta_2\ket{1}$, $\ket{\phi_3}=\text{cos}\theta_3\ket{0}+\text{sin}\theta_3\ket{1}$.
These states are of interest because they can be prepared in experiments starting with an input of  two entangled pairs of qubits~\cite{expts}.  
Previous numerical studies of $|\psi_3\rangle$~\cite{Emary} established upper and lower bounds on $S_{\text{max}}(\psi_3)$ for a given $\tau(\psi_3)$,
\begin{equation}
\label{emmabound}
| \frac{1}{16}S^2_{\text{max}}(\psi_3)-1 | \leq \tau(\psi_3) \leq \frac{1}{32} S_{\text{max}}^2(\psi_3).
\end{equation}
A comparison of Eq.~(\ref{emmabound}) to Eq.~(\ref{gghzresult0}) shows that Eq.~(\ref{gghzresult0}) coincides with the lower bound on $S_{\text{max}}(\psi_3)$. 
Hence, the GGHZ states have the minimum value of $S_{\text{max}}(\psi_3)$ for a given amount of $\tau(\psi_3)$ (Fig. 1). We show below that the MS states, which also belong to this family, can achieve the upper bound and thus give the maximum possible value of  $S_{\text{max}}(\psi_3)$ for a given amount of $\tau(\psi_3)$. 

{\it The MS States}: Consider the first term in Eq.~(\ref{inequality0}) with respect to the MS states $|\psi_s\rangle$ in Eq.~(\ref{slice}),
\begin{widetext}
\begin{eqnarray}
\langle\psi_s|ADC|\psi_s\rangle&=&
\cos\theta_3\cos\theta_a\cos\theta_d\Big\{\cos\theta_3\cos\theta_c+\sin\theta_3\cos\phi_c\sin\theta_c\Big\}
\nonumber\\
&&\,\,\,\,\,+\sin\theta_a\sin\theta_d\Big\{\cos\theta_3\cos\phi_{ad}\cos\theta_c+\sin\theta_3\cos\phi_{adc}\sin\theta_c\Big\}\nonumber\\
& \leq &
\cos\theta_3\cos\theta_a\cos\theta_d(\cos^2\theta_3+\sin^2\theta_3\cos^2\phi_c)^{\frac{1}{2}}+
\sin\theta_a\sin\theta_d(\cos^2\theta_3\cos^2\phi_{ad}+\sin^2\theta_3\cos^2\phi_{adc})^{\frac{1}{2}}\nonumber\\
& \leq &
\Big\{\cos^2\theta_3\cos^2\theta_d (\cos^2\theta_3+\sin^2\theta_3\cos^2\phi_c)
+\sin^2\theta_d (\cos^2\theta_3\cos^2\phi_{ad}+\sin^2\theta_3\cos^2\phi_{adc})\Big\}^{1\over 2}.
\label{sliceinequality1}
\end{eqnarray}
\end{widetext}
The first inequality is obtained by the use of Eq. (\ref{maxtheta}) to maximize the terms in parentheses individually  with respect to $\theta_c$, and the second inequality is obtained by maximizing the first inequality with respect to $\theta_a$. Inserting Eq.~(\ref{sliceinequality1}), (and similar expressions for $\langle AD'C' \rangle, \langle A'D'C \rangle$ and $\langle A'DC' \rangle$), in the inequality in Eq. (\ref{inequality0}) and using the constraint in Eq.~ (\ref{orthogonal}), we find  a turning point of  $S(\psi_s)$ at 
$\phi_{d}-\phi_{d'}= \theta_d=\theta_{d'}=\pi/2$.
Then from Eqs.~(\ref{sliceinequality1}) and~(\ref{inequality0}),
\begin{widetext}
\begin{eqnarray}
S(\psi_s) 
&\leq & 2\Big\{(\cos^2\theta_3\cos^2\phi_{ad}+\sin^2\theta_3\cos^2\phi_{adc})
+(\cos^2\theta_3\cos^2\phi_{ad^{\prime}}+\sin^2\theta_3\cos^2\phi_{ad^{\prime}c^{\prime}})\Big\}^{1\over 2}\nonumber\\
&+& 2\Big\{(\cos^2\theta_3\cos^2\phi_{a^\prime d}+\sin^2\theta_3\cos^2\phi_{a^\prime dc^\prime})
+(\cos^2\theta_3\cos^2\phi_{a^\prime d^\prime}+\sin^2\theta_3\cos^2\phi_{a^\prime d^\prime c})\Big\}^{1\over 2}\nonumber\\
&\leq & 4\Big\{\cos^2\theta_3+2\sin^2\theta_3\Big\}^{1\over 2}= 4{\sqrt{1+\tau(\psi_s)}}\;.
\label{sliceinequality2}
\end{eqnarray}
\end{widetext}
The second inequality in Eq.~(\ref{sliceinequality2}) is obtained from the first by setting $\cos^2\phi_{adc}=\cos^2\phi_{ad^{\prime}c^{\prime}}=\cos^2\phi_{a^\prime dc^\prime}=\cos^2\phi_{a^\prime d^\prime c}=1$, and by noting that since $\phi_{d}-\phi_{d'}=\pi/2$,  $\cos^2\phi_{ad}=\sin^2\phi_{ad'}$ and $\cos^2\phi_{a'd}=\sin^2\phi_{a'd'}$. The final equality  follows from $\tau(\psi_s)=\sin^2\theta_3$, yielding the desired result of Eq.~(\ref{stauslice}) for $S_{\text{max}}(\psi_s)$. The other turning point of $S(\psi_s)$ at $\phi_d - \phi_{d'} =0$ yields a lower value of $S(\psi_s)$, so the expression in Eq.~(\ref{sliceinequality2}) gives the global maximum.
A set of measurement angles which realizes $S_{\text{max}}(\psi_s)$ [ Eq.~(\ref{stauslice})], is 
$\theta_a=\theta_{a'} = \theta_d=\theta_{d'}=\pi/2, \tan\theta_c = \tan\theta_{c'}=\sqrt{2}\tan\theta_3$,  $\phi_{adc}=\phi_{ad'c'}=\phi_{a'd'c}=0$, $\phi_{a'dc'}=\pi$, $\phi_{c'}=-\phi_{c}=\pi/4$, $\phi_d-\phi_{d'}=\pi/2$. Notice that the only difference between these angles and the optimal measurement angles for the GGHZ states in the regime $\tau > 1/3$ is that $\vec c$ and $\vec{c}'$ do not lie in the $xy$-plane.
Comparison of  Eq.~(\ref{stauslice}) to the numerical bounds in Eq.~(\ref{emmabound})~\cite{Emary}, shows that it corresponds to the upper bound on $S_{\text{max}}(\psi_3)$, confirming that this is the maximum possible value of $S(\psi_s)$ as a function of $\tau$. We note that the states obtained by swapping the second and third qubits also yield $S_{\text{max}}$ as in Eq.~(\ref{stauslice}).

From Eq.~(\ref{stauslice}), it is clear that all MS  states can violate the Svetlichny inequality (Fig.~1). Furthermore, we can compare Eq.~(\ref{stauslice}) to the entanglement-nonlocality relationship for 2-qubit pure states $\ket {\phi}$~\cite{Gisin1, Popescu1},
$\text{CHSH}_{\text{max}}(\phi) = 2\sqrt{1+\tau_{12}(\phi)}$.
$\text{CHSH}_{\text{max}}(\phi)$ is the maximum expectation value of the CHSH operator~\cite{CHSH}, and the tangle $\tau_{12}(\phi)={\cal C}^2_{12}(\phi)$ measures the amount of bipartite entanglement in the state~\cite{Wootters,3tangle}. 
Eq.~(\ref{stauslice}) 
for the MS states is directly analogous to this 2-qubit result.

Our analysis shows that the nonlocal correlations in GGHZ states and MS states appear to be quite different although they both belong to the GHZ class. Notice that  for the GGHZ states when $\theta_1=0$ in Eq.~(1), we obtain a 3-qubit product state, whereas for the MS states, setting $\theta_3=0$ in Eq.~(2) yields a product of a maximally entangled state of  2 qubits and the state $\ket{0}$ for the third qubit. 
As $\theta_1$ and $\theta_3$ are increased the GGHZ and MS states both become tripartite entangled, but in different ways due to the different initial states, thereby leading to the differences in nonlocality seen in Eqs.~(3) and (4).

{\it Conclusion}: In summary, we have obtained useful but surprising relationships between tripartite entanglement and nonlocality for the GGHZ and MS states. Previous studies~\cite{Scarani, Zukowski} have found that the GGHZ states do not violate any Bell inequality for $\tau < 1/4$. Here we have shown that the regime of nonviolation is in fact much larger $(\tau < 1/2)$ for the Svetlichny inequality. 
What does the nonviolation of Svetlichny's inequality by some GGHZ states mean? 
Perhaps their nonlocality will be revealed by some other Bell-type inequality, unless one finds an explicit hidden-variable model which reproduces the correlations in these states. 
An interesting topic of further study is the connection between nonlocality and tripartite information in GGHZ states as defined in~\cite{LPW}. Another question of practical interest is the physical significance of the fact that all MS states  violate the Svetlichny inequality and their possible usefulness  for specific information processing tasks. 
In future work, we plan to extend our analysis to W-class states~\cite{GHZClass} 
and more generally, to multipartite nonlocality in an $n$-qubit system via a generalization of the Svetlichny inequality~\cite{NQubit, NSvetlichny}.

 {\it Acknowlegements:} We thank A. Kabra and S. Bandhyopadhyay for helpful discussions. SG was supported by NSERC. PR thanks M. Toshniwal and DST for support.

\end{document}